\documentclass[12pt]{iopart}

\usepackage{iopams} 
\usepackage{siunitx}
\usepackage{graphicx}

\begin{document}

\title{Enhanced oscillation lifetime of a Bose-Einstein condensate in the 3D/1D crossover}

\author{B. Yuen, I. J. M. Barr, J. P. Cotter, E. Butler and E. A. Hinds}

\address{Centre for Cold Matter, Blackett Laboratory, Imperial College London, Prince Consort Road, London SW7 2AZ, United Kingdom}
\ead{ed.hinds@imperial.ac.uk}
\vspace{10pt}
\begin{indented}
\item[]February 2014
\end{indented}

\begin{abstract}
We have measured the damped motion of a trapped Bose-Einstein condensate, oscillating with respect to a thermal cloud. The cigar-shaped trapping potential provides enough transverse confinement that the dynamics of the system are intermediate between three-dimensional and one-dimensional. We find that oscillations persist for longer than expected for a three-dimensional gas. We attribute this to the suppressed occupation of transverse momentum states, which are essential for damping.
\end{abstract}

%
\vspace{2pc}

\section{Introduction}

Trapped, ultracold gases offer a versatile way to investigate quantum many-body physics. Well-isolated from their surroundings, they can be controlled to cover a wide parameter space,  giving access to regimes beyond the reach of other condensed matter experiments \cite{Levin12}.  Confinement reduces the dimensionality of a gas when the atoms have insufficient energy to reach excited quantum levels. For example, pancake-shaped traps can produce a two-dimensional (2D) gas, while a cigar-shaped trap can confine it to one dimension (1D) \cite{Bagnato91}. While the static properties of atomic Bose-Einstein condensates (BEC) are generally well understood \cite{Pethick02} the dynamical behaviour remains an active area of study \cite{Polkovnikov11}. In the early days of atomic BEC, oscillations of the shape were studied, primarily to establish the superfluidity of the condensate, and it was noticed that these oscillations were damped \cite{Jin96, Mewes96} at a  rate that depended strongly on the temperature \cite{Jin97}.  An explanation for this was offered by Landau damping \cite{Liu97, Pitaevskii1997}, in which a low-energy excitation of the condensate is dissipated into the thermal cloud by scattering phonons from lower to higher energy. Fedichev \textit{et al.} \cite{Fedichev98a, Fedichev98b} extended this theory to the case of a trapped gas and showed that the damping is determined predominantly by the condensate boundary region, resulting in a different damping rate from that of a
spatially homogeneous gas. This theory found reasonable agreement with \cite{Jin97}, and similar agreement was found with the measured damping rate of the scissors mode of oscillation \cite{Marago01}.  

Subsequently, Stamper-Kurn \textit{et al.} \cite{StamperKurn98} excited a cigar-shaped condensate to move rigidly along its length, out of phase with its thermal component. They saw that this second-sound motion \cite{Zaremba98} was damped, and noted that collisions neglected in the Landau theory might play a role because the hydrodynamicity -- the thermal cloud collision rate divided by the oscillation frequency -- was not small. The damping of this mode was also noted in \cite{Ferlaino02} and was studied extensively by Meppelink \textit{et al.} \cite{Meppelink09}. They found qualitative agreement with \cite{Fedichev98a} at low values of hydrodynamicity, with a strongly growing discrepancy at higher values, demonstrating the breakdown of the Landau theory at high density. 

Oscillations of long, thin condensates in the 1D regime \cite{Moritz03} have very different behaviour, with no damping \cite{Kinoshita06} unless corrugation is added to the trapping potential \cite{Fertig05}. This raises the question of how the damping evolves from the 3D rate, through the crossover regime where no analytic theory currently exists, to a complete absence of damping in 1D. Oscillation frequencies have been measured in this crossover regime \cite{Kottke05, Fang14}, but not the damping rate. In this article, we measure the damping rate for dipole oscillations of a condensate in the crossover regime as a function of temperature, and compare our results with measurements of \cite{Meppelink09} and the theory of \cite{Fedichev98a, Fedichev98b}. We find that the oscillations in our experiment persist for longer than expected for a 3D gas and propose that this is the consequence of suppressed radial excitations due to the tight transverse confinement of the atoms.

\section{Condensate oscillations in a thermal background}

\begin{figure}
\centering
\includegraphics[width= 0.6 \textwidth]{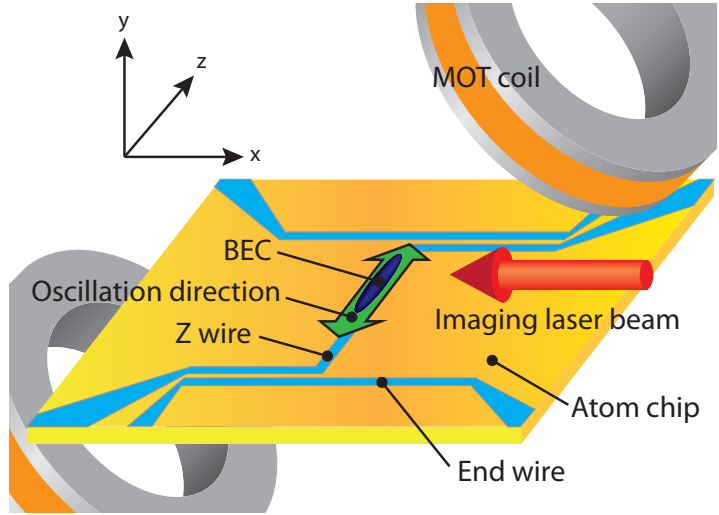}
\caption{Diagram of the apparatus. Cold $^{87}$Rb atoms are delivered from an LVIS source \cite{Sinclair05} to a reflection-magneto-optical trap formed on an atom chip. The atoms are passed to a long, thin magnetic trap formed by a current in the Z-wire together with a uniform bias field along $x$. After evaporative cooling, these form a BEC (dark blue). A brief ac current in the end wire excites the condensate to oscillate along $z$, as indicated by the green arrows. After some time, the cloud is released and allowed to fall for $2\,$ms under gravity along $y$, before being imaged along $x$ by absorption of a laser beam, shown in red. }
\label{fig:apparatus}
\end{figure}

We produce highly elongated, finite temperature condensates \cite{Yuen14} with the apparatus illustrated in figure\,\ref{fig:apparatus}. A magneto-optical trap (MOT) cools and collects $^{87}$Rb atoms a  few millimetres away from the surface of an atom chip \cite{Reichel99}.  The MOT is then turned off, and the atoms are transferred to a Ioffe-Pritchard trap approximately \SI{110}{\micro\metre} from the surface of the chip \cite{Sewell2010, Baumgartner2010}. The magnetic trapping field is produced by current in a Z-shaped wire on the chip, with its central section along $z$, together with an external bias field along $x$. The high magnetic field gradient near the centre of the  Z-wire gives tight radial ($x$, $y$) confinement with a harmonic oscillation frequency of $\omega_{\rho}/2\pi = 1.4~\mathrm{kHz}$. Axial ($z$) confinement  is produced by the currents in the ends of the Z-wire and in the end wires (figure\,\ref{fig:apparatus}), giving an axial frequency of $3~\mathrm{Hz}$.

We cool the trapped gas further by forced evaporation, using an rf field to flip the spins of the most energetic atoms so that they are ejected from the trap \cite{ketterle_evaporative_1996}. By sweeping the escape energy down to a few kilohertz above the  bottom of the trap, we produce an almost pure BEC of approximately $10^4$ atoms at a temperature of $\sim150$\,nK. Minor defects in  the chip wire cause the current to meander slightly from side to side, producing small undulations of the trapping potential that make local minima along the $z$ axis up to a microkelvin in depth \cite{Fortagh02,Jones03}. We adjust the centre of the axial trap so that the BEC forms in one of these, which is harmonic over a small region, with a characteristic frequency of $\omega_z/2\pi = 10~\mathrm{Hz}$. The condensed atoms are confined to that region, while the higher-energy atoms in the thermal component of the gas explore a larger axial range, and experience an anharmonic potential.

When the rf field is turned off, the atoms warm up at approximately $50~\mathrm{nK/s}$, presumably due to noise in the apparatus. To counteract this, we leave the rf field on, so that atoms above some fixed energy are able to leave the trap. Over a few milliseconds the cloud comes to equilibrium at the temperature where the heating is balanced by the evaporative cooling. We select a desired temperature in the range $150-310$~nK by adjusting the rf frequency. The temperature remains fixed over the next  500~ms, while the number of trapped atoms decreases, typically by a few percent.  

\begin{figure*}[t]
\centering
\footnotesize{ \input{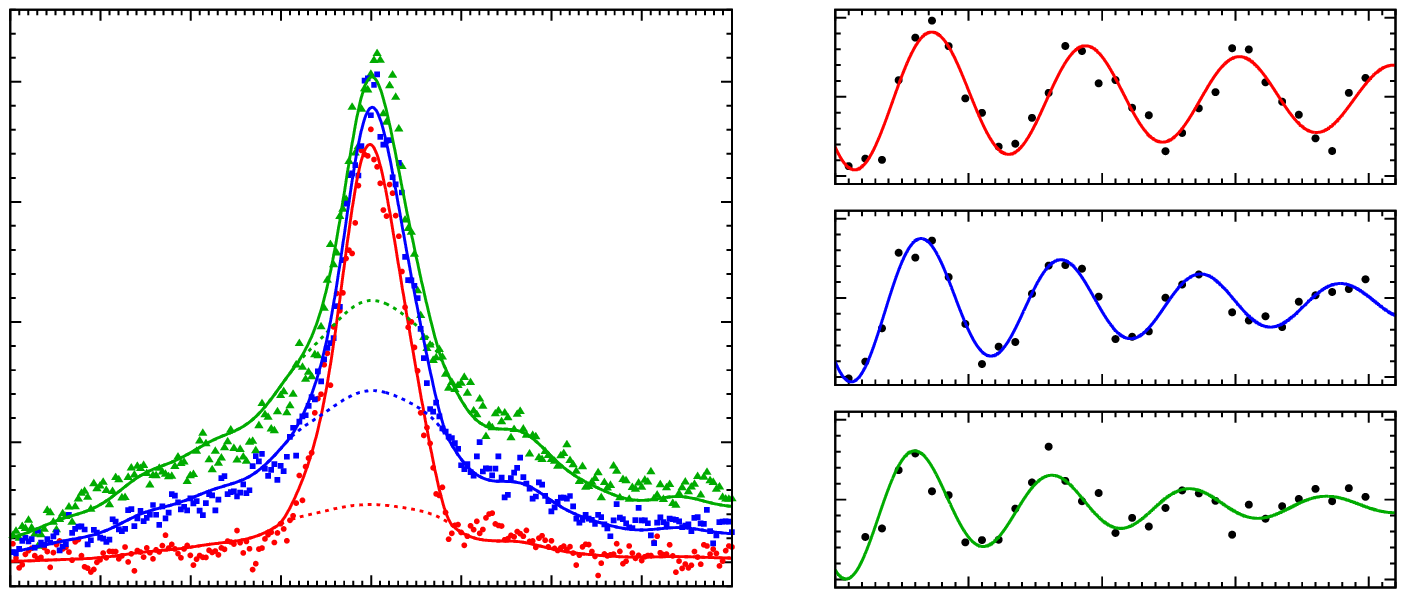} }
\caption{Density profiles of ultracold atom clouds and oscillations of the condensate. (a) Axial column density profiles measured at three temperatures. Red dots: 155(3)\,nK. Blue squares: 251(3)\,nK. Green triangles: 305(3)\,nK. Solid lines: fits using theory described in the text, which takes into account the irregular potential. Dotted lines: profiles of the thermal component of the cloud, determined by the same fit to theory. (b-d) Condensate oscillations for the same three temperatures. Points show the centre of mass of the condensed component after  a period of free oscillation. Lines show the fits to the damped sinusoid in (\ref{eq:dampedSin}). These fits give (b) $\gamma=2.0(6)\,\mbox{s}^{-1}$ at $155$\,nK. (c)  $\gamma=3.8(5)\,\mbox{s}^{-1}$ at $251$\,nK. (d) $\gamma=5.7(1.2)\,\mbox{s}^{-1}$ at $305$\,nK.
}
\label{fig:measurements}
\end{figure*}

Our aim is to observe the oscillation of condensed atoms moving through the thermal cloud in order to determine the damping rate as the system equilibrates. To resonantly excite axial condensate oscillations, we drive an oscillating current in one of the end wires at $10~\mathrm{Hz}$ for two periods. After this time, the condensate's centre of mass is left oscillating with an initial amplitude of \SI{\sim12}{\micro\metre}. The thermal atoms are largely unaffected because they  explore the region outside the local potential minimum and are therefore not resonant with this drive. 

We allow the condensate to oscillate through the thermal background for a time $t$, before switching off the trap and imaging the cloud to determine the condensate's centre of mass. By increasing $t$ in $12.5\,$ms steps over a total of $400\,$ms, we build up a data set of the damped oscillation. We repeat this process for clouds at different temperatures which we influence by setting the frequency of the rf field as described above. Thus, we observe how the system damps as a function of temperature.


\section{Measuring the temperature, condensate centre of mass, and damping rate}

We determine the temperature of the gas, and centre of mass of the condensate from an absorption image.
To image the atom cloud, we release it from the trap (gravity is up in figure\,\ref{fig:apparatus}), wait for 2~ms, illuminate it with resonant laser light and view the absorption along $x$ using a CCD camera. This image is then integrated over $y$ to obtain the one-dimensional axial number density profile of the cloud,  $n(z)= \int \mathrm{d}x \,\mathrm{d}y\,n(\mathbf{r})$. The data points in figure\,\ref{fig:measurements}(a) show axial density profiles measured at three different temperatures. At the lowest temperature (red dots), the atoms are nearly all in the condensate, with very little signal in the broad thermal background, whereas the profile at the highest temperature (green triangles) has a clearly visible thermal population on either side of the cloud. 

Our analysis of the cloud profile builds on the method of \cite{Naraschewski1998}. The trapping potential is well described by $U(\mathbf{r})={1 \over 2} m\omega_{\rho}^2 \rho^2+V(z)$, where $\rho$ is the radial displacement, and $V(z)$ is the potential on axis, including the irregularity caused by the meandering current. We determine $V(z)$ from the axial density distribution of cold, non-condensed clouds as described in \cite{Jones2004}. Knowing $U(\mathbf{r})$, we estimate the number density profile of the condensate, $n_{c}(\mathbf{r})$, using the Thomas-Fermi approximation. The profile of the thermal  component is calculated by integrating the Bose-Einstein distribution over the effective potential $2 g n_{c}(\mathbf{r}) + U(\mathbf{r})$, where the first term is the mean-field energy of thermal atoms inside the condensate. The cloud is then allowed to evolve freely for $2\,$ms to account for the period of free fall (though we find that this makes no significant difference to the axial profile). We fit this theoretical cloud to the measured density profile $n(z)$ in order to determine the temperature, the position of the condensate, and the peak condensate number density $n_{c}(0)$. We note that the Thomas Fermi approximation is not well satisfied in our 3D/1D condensates, but we find from simulations \cite{FuturePaper} that this method still yields accurate temperatures, while the peak condensate density is underestimated, typically by 10\%. These fits, shown in figure\,\ref{fig:measurements}(a) as solid lines, are in excellent agreement with the clouds we observe.  For the three clouds that are plotted in figure\,\ref{fig:measurements}(a), we determined the temperatures $155(3)$, $251(3)$ and $305(3)$\,nK. The dotted lines show the thermal cloud density within the condensed regions.

In our experiments, the temperature fluctuates by less than $10$\,nK from one realisation to the next -- mainly because of fluctuations in the initial number of magnetically trapped atoms -- and  drifts by less than $\pm20$\,nK over an hour. The position of the BEC  is very stable, fluctuating from shot to shot by less than \SI{1}{\micro\metre}, which we associate with mechanical instability of the camera and mirror mounts. It does not drift significantly over an hour.

Figures\,\ref{fig:measurements}(b-d) show plots of the condensate centre of mass oscillations we measure at each of the three temperatures used in figure\,\ref{fig:measurements}(a). It takes approximately 30 minutes to collect the data points for one plot. We have analysed 33 such time sequences, covering a range of temperatures from $150$\,nK up to $310$\,nK. In each case, the motion is well described by the exponentially damped sinusoid,
\begin{equation}
z(t) = A \mathrm{e}^{-\gamma t} \mathrm{sin}\left( \omega t + \phi \right) + C,
\label{eq:dampedSin}
\end{equation}
where $A$, $\gamma$, $\omega$, $\phi$ and $C$ are fit parameters. Parameters $A$, $\phi$ and $C$ are independent of temperature, and $\omega$ increases only slightly (by 10\%) over this range of temperatures. By contrast, $\gamma$ depends significantly on temperature, increasing by a factor of three.

\section{Results and discussion}

\begin{figure}[t]
\centering
\includegraphics[width =0.6 \columnwidth]{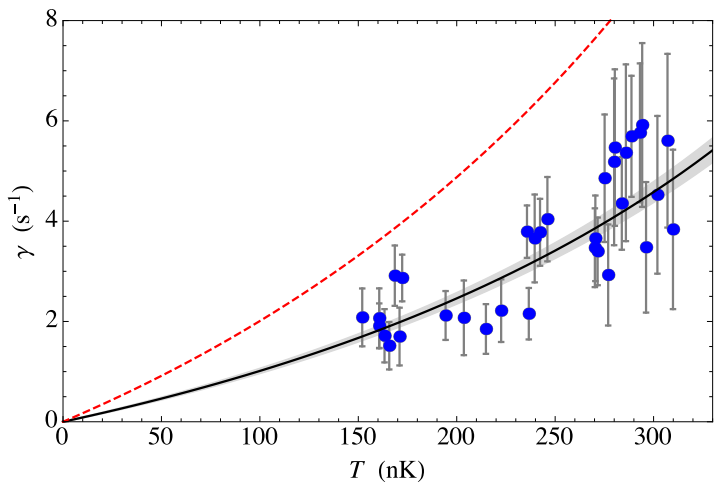}
\caption{Damping rate measured as a function of temperature for the oscillation of our highly elongated BEC. Each point is derived by fitting the oscillation of  32 cloud images to (\ref{eq:dampedSin}).  Vertical error bars show the $1\sigma$ uncertainty in $\gamma$. Horizontal error bars are smaller than the symbol. Solid line: a least squares fit of (\ref{eq:gammatheory}) to our data gives  $A_{\nu}=3.53(15)$. Shading indicates the standard error from the fit. Dashed line: damping rate given by (\ref{eq:gammatheory}) taking $A_{\nu}=7$, as observed with the 3D condensate of \cite{Meppelink09, Straten14}.}
\label{fig:results}
\end{figure}

In figure\,\ref{fig:results} we plot the damping rate $\gamma$ as a function of the temperature, each point being the result of fitting one oscillation curve. The temperature assigned to one point is the mean of the $\sim30$ temperatures measured in that curve, and this has a standard error smaller than the symbols in the plot. A vertical error bar indicates the $1\sigma$ uncertainty in $\gamma$ for each fit.  We note that the dissipated energy of the oscillation has no significant influence on the temperature, because it corresponds to a negligible rise of  $\sim2\,$nK.

Damping rates measured in 3D BEC oscillation experiments have generally been consistent with the Landau damping theory \cite{Jin96, Mewes96, Jin97, Fedichev98a, Fedichev98b, Marago01, Meppelink09}.  For a 3D trapped cloud making small oscillations at a frequency $\omega$ close to the trap frequency, this theory  gives the damping rate as (see equation (18) of Ref.~\cite{Fedichev98a})
\begin{equation} \label{eq:gammatheory}
\Gamma_{\nu} = 
A_{\nu}\, \omega \frac{k_{\mathrm B} T}{g\sqrt{n_c(0)}} 
	 a^{3/2} \,.
\end{equation}
Here $A_{\nu}$ is a numerical coefficient that depends on which collective mode $\nu$ is excited, $a$ is the s-wave scattering length, $n_c(0)$ is the peak number density of the condensate and $g=4 \pi \hbar^2 a/m$ is the usual nonlinear coupling parameter (the $\mu$ of \cite{Fedichev98a} is the same as our $g n_c(0)$).  At each temperature our measurements give values for the number density and oscillation frequency, from which we construct empirical functions  $n_c(0;T)$ and $\omega(T)$. Using these functions, we fit (\ref{eq:gammatheory}) to our data with $A_{\nu}$ as the only free parameter. The result is $A_{\nu}=3.53(15)$. The solid line in figure\,\ref{fig:results} shows this best fit, with the shaded region covering the standard deviation. This theory describes our data well, giving a reduced $\chi^2$ of $0.90$.

The measurements of Meppelink \textit{et al.} in \cite{Meppelink09} involve the same mode as our experiment, but the comparison of their damping at low hydrodynamicity with (\ref{eq:gammatheory}) yields a coefficient $A_{\nu}=7$ \cite{Meppelink09, Straten14}. The essential difference between these two experiments is in the dimensionality of the trapped gas. In \cite{Meppelink09} the chemical potential was at least 28 times higher than the radial excitation energy, placing their cloud firmly in the 3D regime. By contrast, the chemical potential of our cloud is aproximately twice the quantum of radial excitation, which places it in the crossover regime between 3D and 1D. The temperature dependence of our result indicates that the same Landau damping idea still applies, even in this crossover regime, but  the density of states, which enters through the use of Fermi's golden rule to obtain (\ref{eq:gammatheory}), should be modified to account for the quantisation of the radial excitations \cite{Yuen14}. Physically, the thermal excitations in this case are more likely to be along $z$, in which case they cannot contribute to the damping, and $A_{\nu}$ is correspondingly reduced.

\begin{figure}[t]
\centering
\includegraphics[width =0.6 \columnwidth]{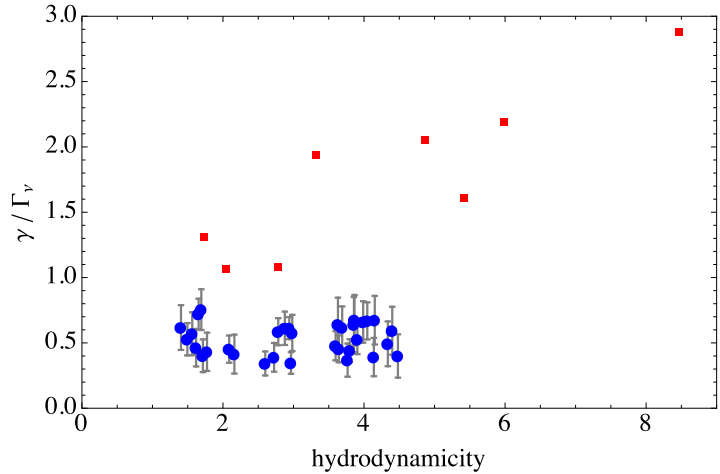}
\caption{Ratio of measured damping rates $\gamma$ to the $\Gamma_{\nu}$ of (\ref{eq:gammatheory}), taking $A_{\nu}=7$, plotted versus hydrodynamicity \cite{Straten14}. Red squares: data of \cite{Meppelink09}, showing an increase in this ratio with increasing hydrodynamicity. Blue circles: our data, showing a suppressed value of $A_{\nu}$ and no excess damping up to a hydrodynamicity of 5.}
\label{fig:hydro}
\end{figure}

The Utrecht experiment \cite{Meppelink09} measured the damping over a wide range of hydrodynamicity. Following in the spirit of \cite{Meppelink09}, the red squares in figure\,\ref{fig:hydro} plot the ratio of their measured damping rates $\gamma$ to the $\Gamma_{\nu}$ of (\ref{eq:gammatheory}), with $A_{\nu}=7$ \cite{Straten14}, plotted versus hydrodynamicity. At low hydrodynamicity, the ratio approaches 1 in their data,  and 0.5 in our data (blue circles), as discussed above. Further, the Utrecht data shows an increase in this ratio as the hydrodynamicity increases, indicating that collisional processes, not  incorporated in the model of (\ref{eq:gammatheory}), play an important role in damping this dipole mode.  Figure\,\ref{fig:hydro} shows that such an increase does not occur in our case. We suggest that this too is a consequence of the discrete radial excitation spectrum, which although broadened at higher collision rates, remains discrete far above a hydrodynamicity of 5 and therefore suppresses the ability of thermal-thermal collisions to contribute to the damping.

Following \cite{Meppelink09}, we have taken the hydrodynamicity in figure\,\ref{fig:hydro} to be $ n_{\mathrm{th}} \langle v_{\mathrm{rel}}\rangle \sigma / \omega_z$, 
where $n_{\mathrm{th}}=N_{\mathrm{th}} m^{3/2} \omega_{\rho}^2 \omega_z / (4 \pi k_B T)^{3/2}$ is the average thermal atom number density experienced by thermal atoms in the harmonic trap, according to the Maxwell Boltzmann distribution. The quantity $\langle v_{\mathrm{rel}}\rangle=4 \left[ k_B T/(\pi m) \right]^{1/2}$ is the mean relative speed between thermal atoms, and $\sigma=8 \pi a^2$ is the s-wave scattering cross-section. In future, it would be better to derive the thermal density from the Bose-Einstein distribution in the harmonic trap, which fixes the  mean density at  $0.55\, \zeta(\frac{3}{2})(2\pi m k_B T)^{3/2}/ h^3$, $\zeta$ being the Reimann zeta function. This makes no difference to our conclusions here, but will be important for any future quantitative study of the corrections to Landau damping. 

In all the damping experiments, the energy in the initial coherent motion is very large compared with  $\hbar \omega$. Indeed, the ratio of these is generally greater than the number of atoms in the cloud. It is therefore interesting that the analytical theory reproduces the measured 3D damping rates, because the theory assumes a  Bogoliubov mode of energy $\hbar \omega$ that is weakly excited.  The agreement between experiment and theory indicates that the damping rate calculated for weak excitations is still applicable when the excitation is strong. 

Collective excitations have been simulated numerically using the method of Zaremba, Nikuni  and Griffin \cite{Zaremba99}, which makes Hartree-Fock and semi-classical approximations to derive a mean field equation for the condensate coupled to a Boltzmann equation for the thermal cloud. Simulations by Jackson and Zaremba \cite{Jackson01, Jackson02, Jackson02b}, have proved to be in good agreement with the 3D experiments \cite{Marago01, Jin97, Chevy02} respectively. 
However, in the 3D/1D cross-over where $\hbar \omega_{\rho} \sim \mu \sim k_B T$, the quantisation of the radial excitations is not well approximated by a semi-classical treatment, as we have shown here. A fully quantum treatment may be possible using the perturbative approach of \cite{Pitaevskii1997, Guilleumas03}, but we are not aware of any such treatment in the 3D/1D crossover regime. Our results provide a point of reference for such simulations. In future we hope to vary the transverse width of our trap in order to elucidate further the damping behaviour in this dimensional crossover region.

\section*{Acknowledgments}
We are indebted to Peter van der Straten,  Eugene Zaremba, Gora Shlyapnikov and Nathan Welch for valuable discussions. We acknowledge the expert technical support of Jon Dyne, Steve Maine, and Valerius Gerulis. This work was supported by the European FP7 project AQUTE, by the UK EPSRC and by the Royal Society. 

\section*{References}
\bibliography{oscillationbibliography}

\providecommand{\newblock}{}
\begin{thebibliography}{10}
\expandafter\ifx\csname url\endcsname\relax
  \def\url#1{{\tt #1}}\fi
\expandafter\ifx\csname urlprefix\endcsname\relax\def\urlprefix{URL }\fi
\providecommand{\eprint}[2][]{\url{#2}}

\bibitem{Levin12}
{Levin} K, {Fetter} A and {Stamper-Kurn} D 2012 {\em Ultracold Bosonic and
  Fermionic Gases\/} ({\em Contemporary Concepts of Condensed Matter Science\/}
  vol~5) (Elsevier)

\bibitem{Bagnato91}
{Bagnato} V and {Kleppner} D 1991 {\em Phys. Rev. A\/} {\bf 44} 7439

\bibitem{Pethick02}
{Pethick} C and {Smith} H 2002 {\em Bose-Einstein condensation in dilute
  gases\/} (Cambridge: Cambridge university press)

\bibitem{Polkovnikov11}
{Polkovnikov} A, {Sengupta} K, {Silva} A and {Vengalattore} M 2011 {\em Rev.
  Mod. Phys.\/} {\bf 83} 863

\bibitem{Jin96}
{Jin} D~S, {Ensher} J~R, {Mathews} M~R, {Wieman} C~E, {Durfee} D~S and
  {Cornell} E~A 1996 {\em Phys. Rev. Lett.\/} {\bf 77} 420

\bibitem{Mewes96}
{Mewes} M~O, {Andrews} M~R, {van Druten} N~J, {Kurn} D~M, {Durfee} D~S and
  {Townsend} C~Gand~{Ketterle} W 1996 {\em Phys. Rev. Lett.\/} {\bf 77} 988

\bibitem{Jin97}
Jin D~S, Matthews M~R, Ensher J~R, Wieman C~E and Cornell E~A 1997 {\em Phys.
  Rev. Lett.\/} {\bf 78}(5) 764

\bibitem{Liu97}
{Liu} W~V 1997 {\em Phys. Rev. Lett.\/} {\bf 79} 4056

\bibitem{Pitaevskii1997}
{Pitaevskii} L~P and {Stringari} S 1997 {\em Phys. Lett. A\/} {\bf 235}
  398--402

\bibitem{Fedichev98a}
{Fedichev} P~O, {Shlyapnikov} G~V and {Walraven} J~T~M 1998 {\em Phys. Rev.
  Lett.\/} {\bf 80} 2269--2272

\bibitem{Fedichev98b}
{Fedichev} P~O and {Shlyapnikov} G~V 1998 {\em \PR A\/} {\bf 58} 3146--3158

\bibitem{Marago01}
{Marag{\`o}} O, {Hechenblaikner} G, {Hodby} E and {Foot} C 2001 {\em Phys. Rev.
  Lett.\/} {\bf 86} 3938

\bibitem{StamperKurn98}
{Stamper-Kurn} D~M, {Miesner} H~J, {Inouye} S, {Andrews} M~R and {Ketterle} W
  1998 {\em Phys. Rev. Lett.\/} {\bf 81} 500--503

\bibitem{Zaremba98}
{Zaremba} E, {Griffin} A and {Nikuni} T 1998 {\em Phys. Rev. A\/} {\bf 57} 4695

\bibitem{Ferlaino02}
{Ferlaino} F, {Maddaloni} P, {Burger} S, {Cataliotti} F~S, {Fort} C, {Modugno}
  M and {Inguscio} M 2002 {\em \PR A\/} {\bf 66} 011604

\bibitem{Meppelink09}
{Meppelink} R, {Koller} S~B, {Vogels} J~M, {Stoof} H~T~C and {van der Straten}
  P 2009 {\em Phys. Rev. Lett.\/} {\bf 103} 265301

\bibitem{Moritz03}
{Moritz} H, {St\"oferle} T, {Kohl} M and {Esslinger} T 2003 {\em Phys. Rev.
  Lett.\/} {\bf 91} 250402

\bibitem{Kinoshita06}
{Kinoshita} T, {Wenger} T and {Weiss} D~S 2006 {\em Nature (London)\/} {\bf
  440} 900--903

\bibitem{Fertig05}
{Fertig} C~D, {O'Hara} K~M, {Huckans} J~H, {Rolston} S~L, {Phillips} W~D and
  {Porto} J~V 2005 {\em \PRL\/} {\bf 94} 120403

\bibitem{Kottke05}
Kottke M, Schulte T, Cacciapuoti L, Hellweg D, Drenkelforth S, Ertmer W and
  Arlt J~J 2005 {\em \PR A\/} {\bf 72}(5) 053631

\bibitem{Fang14}
{Fang} B, {Carleo} G, {Johnson} A and {Bouchoule} I 2014 {\em \PRL\/} {\bf 113}
  035301

\bibitem{Yuen14}
{Yuen} B 2014 {\em Production and Oscillations of a Bose-Einstein Condensate\/}
  Ph.D. thesis Imperial College London

\bibitem{Reichel99}
{Reichel} J, {H{\"a}nsel} W and {H{\"a}nsch} T~W 1999 {\em \PRL\/} {\bf 83}
  3398--3401

\bibitem{Sewell2010}
{Sewell} R~J, {Dingjan} J, {Baumg{\"a}rtner} F, {Llorente-Garc{\'{\i}}a} I,
  {Eriksson} S, {Hinds} E~A, {Lewis} G, {Srinivasan} P, {Moktadir} Z,
  {Gollasch} C~O and {Kraft} M 2010 {\em J. Phys. B\/} {\bf 43} 051003

\bibitem{Baumgartner2010}
{Baumg\"artner} F, {Sewell} R~J, {Eriksson} S, {Llorente-Garc{\'{\i}}a} I,
  Dingjan J, Cotter J~P and Hinds E~A 2010 {\em \PRL\/} {\bf 105}(24) 243003

\bibitem{ketterle_evaporative_1996}
{Ketterle} W and {van Druten} N~J 1996 {\em Adv. At. Mol. Opt. Phys.\/} {\bf
  37} 181--236

\bibitem{Fortagh02}
{Fort{\'a}gh} J, {Ott} H, {Kraft} S, {G{\"u}nther} A and {Zimmermann} C 2002
  {\em \PR A\/} {\bf 66} 041604

\bibitem{Jones03}
Jones M~P~A, Vale C~J, Sahagun D, Hall B~V and Hinds E~A 2003 {\em Phys. Rev.
  Lett.\/} {\bf 91}(8) 080401

\bibitem{Sinclair05}
{Sinclair} C~D~J, {Curtis} E~A, {Garcia} I~L, {Retter} J~A, {Hall} B~V,
  {Eriksson} S, {Sauer} B~E and {Hinds} E~A 2005 {\em \PR A\/} {\bf 72} 031603

\bibitem{Naraschewski1998}
{Naraschewski} M and {Stamper-Kurn} D~M 1998 {\em \PR A\/} {\bf 58} 2423--2426

\bibitem{Jones2004}
{Jones} M~P~A, {Vale} C~J, {Sahagun} D, {Hall} B~V, {Eberlein} C~C, {Sauer}
  B~E, {Furusawa} K, {Richardson} D and {Hinds} E~A 2004 {\em J. Phys. B\/}
  {\bf 37} L15--L20

\bibitem{FuturePaper}
A publication is in preparation.

\bibitem{Straten14}
van~der Straten P 2014 private communication. {The data in \cite{Meppelink09}
  were compared with Eq.(17) of \cite{Fedichev98a}, but the comparison should
  have been with Eq.(18) of \cite{Fedichev98a}. In our figures
  \ref{fig:results} and \ref{fig:hydro} this has been corrected.}

\bibitem{Zaremba99}
Zaremba E, Nikuni T and Griffin A 1999 {\em Journal of Low Temperature
  Physics\/} {\bf 116} 277--345

\bibitem{Jackson01}
Jackson B and Zaremba E 2001 {\em Phys. Rev. Lett.\/} {\bf 87}(10) 100404

\bibitem{Jackson02}
{Jackson} B and {Zaremba} E 2002 {\em Phys. Rev. Lett.\/} {\bf 88} 180402

\bibitem{Jackson02b}
Jackson B and Zaremba E 2002 {\em Phys. Rev. Lett.\/} {\bf 89}(15) 150402

\bibitem{Chevy02}
{Chevy} F, {Bretin} V, {Rosenbusch} P, {Madison} K~W and {Dalibard} J 2002 {\em
  Phys. Rev. Lett.\/} {\bf 88} 250402

\bibitem{Guilleumas03}
Guilleumas M and Pitaevskii L~P 2003 {\em Phys. Rev. A\/} {\bf 67}(5) 053607

\end{thebibliography}

\end{document}